# One-dimensional time-Floquet photonic crystal

Neng Wang and Guo Ping Wang*

*Institute of Microscale Optoelectronics, Shenzhen University, Shenzhen, China*

Corresponding author: *gpwang@szu.edu.cn

**Abstract**:

Using the Floquet Hamiltonian derived based on the time-dependent perturbation theory, we investigated the quasienergy bands of a one-dimensional time-Floquet photonic crystal. The time-Floquet photonic crystal contains two alternating layers labeled as A and B, and the permittivity of A layer is modulated periodically in time. We showed that the quasienergy bands are reciprocal when the modulation function is a function of time only, while the quasienergy bands could be nonreciprocal when the permittivity is modulated in both time and space through an unique combination. In the former case, the coupling between the positive (negative) and positive (negative) bands results in quasienergy gaps, while the coupling between the positive and negative bands leads to pairs of exception points, when the modulation is on the real part of the permittivity. In the latter case, the coupling between the positive (negative) and positive (negative) bands still results in quasienergy gaps. However, the coupling between the positive and negative bands leads to quasienergy gaps at a small modulation speed and pairs of exceptional points at a high modulation speed.

## I. Introduction

Time-Floquet systems with parameters periodically modulated in time provide a versatile platform to realize various exotic physical phenomena in both quantum [1-5] and classical physics [6-16], thus have attracted a growing interest in recent years. Periodically modulating the parameters in time breaks the time-reversal symmetry, enabling the time-Floquet system to support topologically nontrivial states that are topologically trivial in the static case, i.e., the Floquet topological states [17, 18], and to possess nonreciprocal band structures [19, 20]. Recently, the Floquet topological states in various classical wave systems were studied [21-25] and experimentally observed [26-30], and several types of nonreciprocal devices based on the time-modulated elements were designed [31-33]. The periodic time-modulation also brings about the discrete translational symmetry in time, which makes the frequency can be converted through adding or subtracting an integer times of the modulation frequency [34], in analogy to the wave vector conversion of the classical waves by a lattice structure. On the other hand, the time modulation requires an external driving in principle, which could make the total energy of the system not necessarily conservative thus introduce non-Hermiticity to the system [35-38], even though there is no gain or loss material used. The coexistence of the non-reciprocity, frequency conversion and non-Hermiticity in the time-Floquet system enables the realization of various attractive applications, such as the broadband nonreciprocal wave amplification [39-42] and frequency-selective wave filtering [43].

In our previous work [35], we have studied the quasienergy bands of a homogeneous medium with a time-periodic complex permitttivity. Here, we extended the study to the case that the periodic time-modulation is applied to the one-dimensional (1D) photonic crystal (PC) with two alternating layers A and B. While the permittivity of B layer is static, the permittivity of A layer is time-modulated as $\varepsilon_a + \delta(z,t)$, where $\varepsilon_a$ is the static part and $\delta(z,t)$ is the time-periodic modulation function of variables time $t$ and spatial coordinate $z$. We calculated and analyzed the quasienergy bands using the Floquet Hamiltonian which is derived from the Maxwell's equations based on the time-dependent perturbation theory. In the vanishing $\delta$ limit, the bands of the 1D PC in the static case copy themselves and shift up and down in the quasienergy space by $n\Omega$ to generate the bands of order $n$, where $n$ is an integer and $\Omega$ is the modulation frequency. Bands of different orders cross and form diabolic points. Here, we call the bands with positive frequencies in the static case (or bands of zeroth order) and their non-zeroth order counterparts positive bands, and the bands with negative frequencies in the static case and their non-zeroth order counterparts negative bands. When $\delta(z,t)$ is a function of $t$ only, namely $\delta(z,t)=\delta g(t)$, the quasienergy bands are proved to be reciprocal. For real and finite $\delta$, the diabolic point formed by two positive (negative) bands are lifted to open a quasienergy gap, while the diabolic point formed by a positive and a negative bands spawns into a pair of exceptional points (EPs). In particular, the EPs formed by band $n$ of $0^{\text{th}}$ ($-1^{\text{st}}$) order and band $-n$ of $1^{\text{st}}$ ($0^{\text{th}}$) order are located at $\Omega/2$ ($-\Omega/2$) in the quasienergy space, similar to the case of a homogeneous medium with time-periodic real permittivity [34].

When $t$ and $z$ are not separable variables in the modulation function, the quasienergy bands could be nonreciprocal. We investigated the simplest case that $\delta(z,t)=\delta\cos(\Omega t-\beta z)$, where $\beta$ is the spatial modulation frequency. We showed that the quasienergy bands are nonreciprocal when $\beta\neq 0$. The diabolic points formed by two positive (negative) bands are still lifted to open quasienergy gaps when $\delta$ is real. However, the diabolic point formed by a positive and a negative bands is lifted to open a quasienergy gap when the modulation frequency $c_m=\Omega/\beta$ is small, but spawns into a pair of EPs when $c_m$ is large for a real $\delta$, which is again similar to the case of a time-modulated homogeneous medium [35].

## II. Formulation of the Floquet Hamiltonian

The unit cell of the 1D time-Floquet PC we consider is schematically shown in Fig. 1, which contains two dielectric layers. The static relative permittivities of the A and B layers are $\varepsilon_a$ and $\varepsilon_b$ , respectively. Consider the perimittivity of the A layer is modulated periodically in time, the time-dependent relative permittivity of the PC is expressed as

$$\varepsilon(z,t)=\varepsilon_s(z)+\varepsilon_m(z,t)=\begin{cases}\varepsilon_a+\delta(z,t), & n\Lambda\le z\le n\Lambda+d_a\\ \varepsilon_b, & n\Lambda+d_a\le z\le(n+1)\Lambda\end{cases}, \tag{1}$$

where $d_a, d_b$ are the thickness of the A and B layers of the unit cell, $\Lambda = d_a + d_b$ is the lattice constant of the PC, and $\delta(z,t)=\delta(z,t+T)$ with $T$ being the modulation period. We first consider the modulation function $\delta(z,t)$ is constant of *z*. For the simplest case, we let $\delta(z,t)=\delta\cos(\Omega t)$, where $\Omega = 2\pi/T$ .

For the sake of mathematical simplicity, we set the permittivity $\varepsilon_0$ , permeability $\mu_0$ and light speed *c* in vacuum as unity. Also we consider all the materials are nonmagnetic, namely the relative permeability is 1. Without loss of generality, we let the electric field to be polarized along the *x* direction. According to the Maxwell's equations, we can obtain the Schrodinger-like equation as

$$\vec{A}\vec{\psi}=i\partial_t[(\vec{B}_0+\vec{B}_t)\vec{\psi}], \tag{2}$$

where

$$\vec{A}=\begin{pmatrix}0 & -i\partial_z\\ -i\partial_z & 0\end{pmatrix},\quad \vec{B}_0=\begin{pmatrix}\varepsilon_s(z) & 0\\ 0 & 1\end{pmatrix},\quad \vec{B}_t=\begin{pmatrix}\varepsilon_m(z,t) & 0\\ 0 & 0\end{pmatrix},\quad \vec{\psi}=\begin{pmatrix}E_x\\ H_y\end{pmatrix}. \tag{3}$$

For the Bloch wavenumber *q*, the time-Floquet solution can be written as $\vec{\psi}=e^{iqz-iQt}\vec{\phi}_q(z,t)$, where $Q$ is the quasi-energy and $\vec{\phi}_q(z,t)=\vec{\phi}_q(z,t+T)$ is the time-periodic function. We use the eigenmodes of the static bands ( $\delta=0$ ) to expand $\vec{\phi}_q$ as

$$\vec{\phi}_q(z,t)=\sum_{j=-\infty}^{\infty}\sum_{m=-\infty}^{\infty}c_{jm}\left|j\right\rangle_q e^{im\Omega t}=\sum_{j=-\infty}^{\infty}\sum_{m=-\infty}^{\infty}c_{jm}\left|j,m\right\rangle_q, \tag{4}$$

where $c_{jm}$ is the expansion coefficient, $\left|j\right\rangle_q$ is the normalized eigenmode of the $j$th band in the static case, $m$ is the order of the frequency and $\left|j,m\right\rangle_q=\left|j\right\rangle_q e^{im\Omega t}$. Substituting Eq. (4) into Eq. (2), we obtain

$$\vec{A}e^{iqz-iQt}\vec{\phi}_q=i\partial_t[(\vec{B}_0+\vec{B}_t)e^{iqz-iQt}\vec{\phi}_q]=e^{iqz-iQt}i\partial_t[(\vec{B}_0+\vec{B}_t)\vec{\phi}_q]+Qe^{iqz-iQt}(\vec{B}_0+\vec{B}_t)\vec{\phi}_q. \tag{5}$$

Eliminating $e^{iqz-iQt}$ on both sides and using Eq. (4), Eq. (5) becomes

$$\begin{aligned}&\vec{A}_q\sum_{j=-\infty}^{\infty}\sum_{m=-\infty}^{\infty}c_{jm}\left|j\right\rangle_q e^{im\Omega t}-i\partial_t[(\vec{B}_0+\vec{B}_v e^{i\Omega t}+\vec{B}_v e^{-i\Omega t})\sum_{j=-\infty}^{\infty}\sum_{m=-\infty}^{\infty}c_{jm}\left|j\right\rangle_q e^{im\Omega t}]\\&=Q(\vec{B}_0+\vec{B}_v e^{i\Omega t}+\vec{B}_v e^{-i\Omega t})\sum_{j=-\infty}^{\infty}\sum_{m=-\infty}^{\infty}c_{jm}\left|j\right\rangle_q e^{im\Omega t},\end{aligned} \tag{6}$$

where

$$\vec{A}_q=e^{-iqz+iQt}\vec{A}(e^{iqz-iQt}),\quad \vec{B}_v=\begin{pmatrix}\varepsilon_m(z) & 0\\ 0 & 0\end{pmatrix}. \tag{7}$$

Inner product $\left\langle j,n\right|_q=\left\langle j\right|_q e^{-in\Omega t}$ from left on both sides and using the orthonormality of the static eigenmodes ($\left\langle l\middle|\vec{A}_q j\right\rangle=\omega_j\delta_{jl}$, $\left\langle l\middle|\vec{B}_0 j\right\rangle=\delta_{jl}$), we obtain

$$\begin{aligned}&\sum_{j=-\infty}^{\infty}\sum_{m=-\infty}^{\infty}c_{jm}[(\omega_j+n\Omega)\delta_{jl}\delta_{mn}+n\Omega(\delta_{mn+1}+\delta_{mn-1})\Delta_{lj}]\\&=Q\sum_{j=-\infty}^{\infty}\sum_{m=-\infty}^{\infty}c_{jm}[\delta_{jl}\delta_{mn}+\delta_{mn+1}+\delta_{mn-1})\Delta_{lj}],\end{aligned} \tag{8}$$

where $\omega_j$ is the eigenfrequency of the $j$th band in the static case, $\delta_{ij}$ is the Kronecker delta function, and $\Delta_{lj}=\left\langle l\middle|\vec{B}_v j\right\rangle$. In the matrix form, Eq. (8) is rewritten as

$$\vec{H}_1\vec{\varphi}=Q\vec{H}_2\vec{\varphi}, \tag{9}$$

where $\vec{\varphi}=(\ldots,c_{j,-1},c_{j,0},c_{j,1},\ldots)^T$, and

$$\overleftrightarrow{H}_1 = \begin{pmatrix} \ddots & & & & & \\ -\Omega\overleftrightarrow{\Delta} & \overleftrightarrow{H}_0 - \Omega\overleftrightarrow{I} & -\Omega\overleftrightarrow{\Delta} & & & \\ & 0 & \overleftrightarrow{H}_0 & 0 & & \\ & & \Omega\overleftrightarrow{\Delta} & \overleftrightarrow{H}_0 + \Omega\overleftrightarrow{I} & \Omega\overleftrightarrow{\Delta} & \\ & & & & \ddots & \end{pmatrix},$$

$$\overleftrightarrow{H}_2 = \begin{pmatrix} \ddots & & & & \\ \overleftrightarrow{\Delta} & \overleftrightarrow{I} & \overleftrightarrow{\Delta} & & \\ & \overleftrightarrow{\Delta} & \overleftrightarrow{I} & \overleftrightarrow{\Delta} & \\ & & \overleftrightarrow{\Delta} & \overleftrightarrow{I} & \overleftrightarrow{\Delta} \\ & & & & \ddots \end{pmatrix}. \quad (10)$$

In Eq. (10), $H_{0,ij} = \omega_j \delta_{ij}$ and $I$ is the identity matrix. Then the quasi-energy $Q$ is obtained as the eigenvalues of the Floquet Hamiltonian $\overleftrightarrow{H}_F = \overleftrightarrow{H}_2^{-1} \cdot \overleftrightarrow{H}_1$.

In the static case, the bands are reciprocal, namely $\omega_j(q) = \omega_j(-q)$. If we choose the center of the A or B layer as the origin, then the static PC possesses the inversion symmetry, which leads $\varepsilon(z) = \varepsilon(-z)$. Assume the normalized electromagnetic fields of the static mode $(\omega, q)$ are $\widetilde{E}_x(z), \widetilde{H}_y(z)$, and note that the electric field is a vector while the magnetic field is a pseudovector, the normalized electromagnetic fields of the static mode $(\omega, -q)$ are then obtained as $\hat{O}_I \widetilde{E}_x(z) = -\widetilde{E}_x(-z), \hat{O}_I \widetilde{H}_y(z) = \widetilde{H}_y(-z)$, where $\hat{O}_I$ is the inversion operator. Then $\Delta_{lj}$ at $-q$ is calculated as

$$\begin{aligned} \Delta_{jl}(-q) &= \delta \int_0^{\Lambda} [-\widetilde{E}_x^{(l)*}(-z)][-\widetilde{E}_x^{(j)}(-z)]\varepsilon(z)dz \\ &= -\delta \int_0^{-\Lambda} \widetilde{E}_x^{(l)*}(-z)\widetilde{E}_x^{(j)}(-z)\varepsilon(-z)d(-z) \\ &= \delta \int_0^{\Lambda} \widetilde{E}_x^{(l)*}(z')e^{iq\Lambda}\widetilde{E}_x^{(j)}(z')e^{-iq\Lambda}\varepsilon(z')dz' = \Delta_{jl}(q). \end{aligned} \quad (11)$$

In Eq. (11), we have used the Bloch theorem $E_x(z+\Lambda) = E_x(z)e^{iq\Lambda}$. Therefore, when the wavenumber is changed from $q$ to $-q$, the block matrices $\overleftrightarrow{H}_0$ and $\overleftrightarrow{\Delta}$ are invariant, and so is the Floquet Hamiltonian $\ddot{H}_F$. This guarantees the quasienergy bands to be reciprocal, i.e. $Q(q) = Q(-q)$.

## III. **Coupling between the positive and positive bands**

Consider that $\varepsilon_a = 1, \varepsilon_b = 100, d_a = 0.9\Lambda, d_b = 0.1\Lambda.$ The static band dispersion and the corresponding eigenmodes of the photonic crystal can be calculated using the transfer matrix method [44], see details in the appendix. The band diagram in the static case is shown in Fig. 2(a), where the positive ( $\omega > 0$ ) and negative ( $\omega < 0$ ) bands are represented by the black and red solid lines, respectively. The band index is indicated on each band. Note that the bands $\pm n$ are mirror symmetric to each other with respect to the zero frequency line.

When the time-modulation is introduced, in the limit of vanishing $\delta$, the Floquet bands can be obtained by copying the static bands and shifting them up and down in the quasienergy space by integer multiples of $\Omega$. Here, we call the bands obtained by shifting the static bands up (down) by $n\Omega$ the bands of $n$th (-$n$th) order. And we call the bands from shifting the positive (negative) static bands also the positive (negative) bands. As shown in Fig. 2(b), only the $0^{th}$, $1^{st}$ and $-1^{st}$ order bands are shown, which are represented by the black, red and blue solid lines, respectively, and the positive and negative bands are represented by the solid and dashed lines. Because the quasienergy bands are reciprocal, we only plotted the bands for $q > 0$. The bands of different orders can cross each other and form diabolic points which are noted as A-F in Fig. 2(b). Points A, B, D, and E are formed by bands differing by an order 1, while points C and F are formed by bands differing by an order 2.

For a finite $\delta$, the Floquet bands can be numerically calculated using Eq. (9) by truncating the band order and the band index to finite numbers. Because of the nonzero matrix $\Delta$, there are couplings between the bands of different orders, which will lift the diabolic points to generate quasienergy gaps, EPs or a mixture of the two. From Eqs. (9) and (10), we can find that the coupling are vanishing small when the order difference of the two bands is greater than 1.

According to Eqs. (9) and (10), when $\delta$ is small, the quasienergy band in the vicinity of the diabolic point formed by bands $|j,1\rangle$ and $|l,0\rangle$ (point A for example) can be well described by the following $2\times 2$ reduced Hamiltonian

$$\vec{H}_{red} = \begin{pmatrix} 1 & \Delta_{lj} \\ \Delta_{jl} & 1 \end{pmatrix}^{-1} \begin{pmatrix} \omega_l & 0 \\ \Omega\Delta_{jl} & \omega_j + \Omega \end{pmatrix} = \frac{1}{1-\Delta_{jl}\Delta_{lj}} \begin{pmatrix} \omega_l - \Omega\Delta_{jl}\Delta_{lj} & -\Delta_{lj}(\omega_j + \Omega) \\ -\Delta_{jl}(\omega_l - \Omega) & \omega_j + \Omega \end{pmatrix}. \quad (12)$$

Note that $\omega_j + \Omega \approx \omega_l$, the quasienergies determined by $\vec{H}_{red}$ are calculated as

$$Q_{\pm} = \frac{1}{2(1-\Delta_{lj}\Delta_{jl})}\left[2\omega_l + \Delta_{lj}\Delta_{jl}(\omega_j - \omega_l) \pm \sqrt{\Delta_{lj}\Delta_{jl}}\sqrt{\Delta_{lj}\Delta_{jl}(\omega_j - \omega_l)^2 + 4\omega_l\omega_j}\right] \quad (13)$$

Since $|\Delta_{lj}\Delta_{jl}| << 1$ when $\delta$ is small,

$$Q_{\pm} \approx \omega_l \pm \sqrt{\Delta_{jl}\Delta_{lj}}\sqrt{\omega_l\omega_j}. \quad (14)$$

When $\omega_l$ and $\omega_j$ have the same sign, $\sqrt{\omega_l\omega_j}$ is real. For pure real $\delta$, $\vec{B}_v$ is Hermitian, then $\Delta_{lj} = \langle l|\vec{B}_v j\rangle = \langle \vec{B}_v l|j\rangle = \langle j|\vec{B}_v l\rangle^* = \Delta^*_{jl}$. Thus $\sqrt{\Delta_{jl}\Delta_{lj}} = |\Delta_{lj}|$ is positive real. Therefore, there is a quasienergy gap with gap size given by

$$\delta Q = Q_+ - Q_- \approx 2|\Delta_{lj}|\sqrt{\omega_l\omega_j}. \quad (15)$$

In Fig. 3(a), we plotted the quasienergy bands evolved from the bands $|2,0\rangle, |2,1\rangle, |3,-1\rangle, |3,0\rangle$, when the time-modulated permittivity becomes $\delta = 0.1$ and the modulation frequency fulfills $\Omega\Lambda/c = 0.5$. As we can see, the diabolic points A and B are lifted and quasienergy gaps open. At point A, we numerically obtain $\omega_2\Lambda/c \approx 3.06, \omega_3\Lambda/c \approx 3.56, \Delta_{23} \approx -1.91\times 10^{-2} + 0.587\times 10^{-2} i$ using Eqs. (A1-A7). According to Eq. (15), the gap size lifted from point A is approximated as $\delta Q\Lambda/c \approx 0.12$, which is very close to the result 0.11 obtained from the full Floquet Hamiltonian $\vec{H}_F$ see Fig. 3(a). There is also a band gap opening at the point C, as

shown in Fig. 3(b). Because the coupling between the bands differing by an order 2 is much weaker, the gap size is several orders smaller.

When $\delta$ is pure imaginary, $\vec{\vec{B}}_v$ is anti-Hermitian, leading to $\Delta_{lj} = \langle l | \vec{\vec{B}}_v j \rangle = -\langle \vec{\vec{B}}_v l | j \rangle = -\langle j | \vec{\vec{B}}_v l \rangle^* = -\Delta^*_{jl}$ . Therefore, the quasienergies in the vicinity of the diabolic point become a complex conjugate pair in between a pair of EPs. According to Eq. (13), the EPs are located at

$$\Delta_{lj}\Delta_{jl}(\omega_j - \omega_l)^2 + 4\omega_l\omega_j = -|\Delta_{jl}|^2 (\omega_j - \omega_l)^2 + 4\omega_l\omega_j = 0. \tag{16}$$

Fig. 4 shows the quasienergy bands near the diabolic points A and B when $\delta = 0.1i$ . The bands near the diabolic points coalesce and form a pair of EPs, which are noted as ($K_1$, $K_2$) and ($K_3$, $K_4$). From Eq. (16) we see that the wavenumbers for the pair of EPs depend on the frequencies $\omega_l, \omega_j$ and the $\Delta_{lj}$. Thus although the diabolic points A and B are corresponding to the same wavenumber, the wavenumbers of the two pairs of EPs spawned from the two diabolic points are different, as shown in Fig. 4(b).

## IV. Coupling between the positive and negative bands

When $\omega_l$ and $\omega_j$ have opposite signs, $\sqrt{\omega_l\omega_j}$ becomes pure imaginary. If $\delta$ is pure real, since $\Delta_{lj} = \Delta^*_{jl}$ , according to Eq. (13), the quasienergies become a complex conjugate pair in between a pair of EPs.

In Fig. 5, we showed the quasienergy bands in the vicinity of the diabolic point D by the black circles when $\delta = 0.2$ and $\Omega\Lambda / c = 0.5$ . A pair of EPs form at $Q = \Omega / 2$ . This is very similar to the case of a time-modulated homogeneous medium [34]. Since $\omega_j = -\omega_l$ and $|\Delta_{1,-1}| << 1$ , according to Eq. (12), the quasienergies near the point D are given by

$$Q_\pm \approx \frac{\Omega}{2} \pm \frac{i}{2}\sqrt{|\Delta_{1-1}|^2 \Omega^2 - (\Omega - 2\omega_1)^2}. \tag{17}$$

Therefore, the EPs are located near $\omega_1 = \frac{1}{2}\Omega(1 \pm |\Delta_{1-1}|)$ . Using Eqs. (A1-A7), we obtained $|\Delta_{1\text{-}1}| \approx 0.0076$ . Then approximately, the wavenumbers for the two EPs are corresponding to $2\omega_1/\Omega = 0.9924$ and $2\omega_1/\Omega = 1.0076$ of the static band, which are marked by the red arrows in Fig. 5(a) and very close to the wavenumbers of the two EPs. The positive (negative) value of the imaginary part of the quasienergy denotes the decay (amplification) rate of the fields in time [34]. From Eq. (17), we see that the maxim imaginary part of the quasienergy is about $\frac{i}{2}|\Delta_{1\text{-}1}|\Omega \approx 0.0019ic/\Lambda$ , agree with the results as shown in Fig. 5(b).

When the modulation frequency is $\Omega\Lambda/c = 4$ , $|3,0\rangle$ and $|-1,1\rangle$ intersects, see Fig. 6(a). If $\delta$ is pure real, the bands near the diabolic point also coalesce and form a pair of EPs, as shown in Figs. 6(b) and 6(c). However, the EPs are no longer fixed to $Q = \pm\Omega/2$ . According to Eq. (13), the locations of the EPs are determined by

$$|\Delta_{3-1}|^2 (\omega_3 - \omega_{-1})^2 + 4\omega_3\omega_{-1} = |\Delta_{3-1}|^2 (\omega_3 + \omega_1)^2 - 4\omega_3\omega_1 = 0. \qquad (18)$$

When $\delta$ is pure imaginary, it is easy to know that the diabolic points formed by the bands differing by an order 1 will be lifted to open quasienergy gaps.

## V. Nonreciprocal band structures

As discussed previously, the quasienergy bands are reciprocal when the time-modulation function is in the form of $\delta(z,t) = \delta g(t)$ . To achieve nonreciprocal band structures, we now consider that $\delta(z,t)$ depends on both $t$ and $z$. For the simplest case, we assume $\delta(z,t) = \delta\cos(\Omega t - \beta z)$ , where $\beta$ is the spatial modulation frequency. Then the modulation speed is defined as $c_m = \Omega/\beta$ , which is in units of $c$. We define a new variable $u = t - z/c_m$ , and use the following transformations,

$$(z,u) = (z, t - z/c_m), \quad (\partial_z, \partial_t) = (\partial_z - \partial_u/c_m, \partial_u). \qquad (19)$$

The time-Floquet solution is then expressed as

$$\tilde{\psi}_q = e^{iqz-iQt}\tilde{\phi}_q = e^{iqz-iQt}\sum_{j=-\infty}^{\infty}\sum_{m=-\infty}^{\infty} c_{jm}\left|j,m\right\rangle_q^{(u)} = e^{i\tilde{q}z-iQu}\sum_{j=-\infty}^{\infty}\sum_{m=-\infty}^{\infty} c_{jm}\left|j\right\rangle_q e^{im\Omega u}, \tag{20}$$

where $\tilde{q} = q - q' = q - Q/c_m$. Substituting Eqs. (19) and (20) into Eq. (2), we arrive at

$$\vec{A}\tilde{\psi} + \vec{C}\tilde{\psi} = i\partial_u[(\vec{B}_0 + \vec{B}_t)\tilde{\psi}], \tag{21}$$

where $\vec{A}, \vec{B}_0, \vec{B}_t$ are defined in Eq. (3), and

$$\vec{C} = \frac{1}{c_m}\begin{pmatrix} 0 & i\partial_u \\ i\partial_u & 0 \end{pmatrix}. \tag{22}$$

Then Eq. (21) is rewritten as

$$\vec{A}e^{i\tilde{q}z-iQu}\tilde{\phi}_q + \vec{C}e^{i\tilde{q}z-iQu}\tilde{\phi}_q = e^{i\tilde{q}z-iQu}Q[(\vec{B}_0 + \vec{B}_t)\tilde{\phi}_q] + e^{i\tilde{q}z-iQu}i\partial_u[(\vec{B}_0 + \vec{B}_t)\tilde{\phi}_q]. \tag{23}$$

Multiply $e^{-i\tilde{q}z+iQu}$ and inner product $\left\langle l,n\right|_q^u = \left\langle l\right|_q e^{-in\Omega u}$ from left, and note that

$$\begin{aligned}
&\left\langle l,n\right|_q^u e^{-i\tilde{q}z}\vec{A}e^{i\tilde{q}z}\left|j,m\right\rangle_q^u = \left\langle l,n\right|_q^u e^{-iqz}e^{iq'z}\vec{A}e^{-iq'z}e^{iqz}\left|j,m\right\rangle_q^u \\
&= \left\langle l\right|_q e^{-iqz}\vec{A}e^{iqz}\left|j\right\rangle_q \delta_{mn} - \frac{Q}{c_m}\left\langle l\right|_q \sigma_x\left|j\right\rangle_q \delta_{mn} = \omega_j\delta_{mn}\delta_{jl} - \frac{Q}{c_m}\Gamma_{lj}\delta_{mn}, \\
&\left\langle l,n\right|_q^u e^{-i\tilde{q}z+iQu}\vec{C}e^{i\tilde{q}z-iQu}\left|j,m\right\rangle_q^u = \frac{Q}{c_m}\left\langle l\right|_q \sigma_x\left|j\right\rangle_q \delta_{mn} - m\beta\delta_{mn}\left\langle l\right|_q \sigma_x\left|j\right\rangle_q \\
&= (\frac{Q}{c_m} - m\beta)\Gamma_{lj}\delta_{mn},
\end{aligned} \tag{24}$$

where

$$\sigma_x = \begin{pmatrix} 0 & 1 \\ 1 & 0 \end{pmatrix}$$

is the Pauli matrix, then we obtain

$$\begin{aligned}
&\sum_{j=-\infty}^{\infty}\sum_{m=-\infty}^{\infty} c_{jm}[\omega_j\delta_{jl}\delta_{mn} - n\beta\delta_{mn}\Gamma_{lj} + n\Omega(\delta_{jl}\delta_{mn} + \delta_{mn+1}\Delta_{lj} + \delta_{mn-1}\Delta_{lj})] \\
&= Q\sum_{j=-\infty}^{\infty}\sum_{m=-\infty}^{\infty} c_{jm}(\delta_{jl}\delta_{mn} + \delta_{mn+1}\Delta_{lj} + \delta_{mn-1}\Delta_{lj}).
\end{aligned} \tag{25}$$

In a matrix form, Eq. (25) is written as

$$\tilde{H}_1\vec{\varphi} = Q\vec{H}_2\vec{\varphi}, \tag{26}$$

where $\vec{H}_2$ is defined in Eq. (10), and

$$\tilde{H}_1 = \begin{pmatrix} \ddots & & & & \\ -\Omega\vec{\Delta} & \vec{H}_0 + \beta\vec{\Gamma} - \Omega\vec{I} & -\Omega\vec{\Delta} & & \\ & & \vec{H}_0 & & \\ & & \Omega\vec{\Delta} & \vec{H}_0 - \beta\vec{\Gamma} + \Omega\vec{I} & \Omega\vec{\Delta} \\ & & & & \ddots \end{pmatrix}.$$

(27)

The quasienergies are obtained as the eigenvalues of the Floquet Hamiltonian $\tilde{H}_F = \vec{H}_2^{-1} \cdot \tilde{H}_1$.

Similarly, the normalized electromagnetic fields of the static modes $(\omega, q)$ and $(\omega, -q)$ can be given by $\tilde{E}_x(z), \tilde{H}_y(z)$ and $\hat{O}_I \tilde{E}_x(z) = -\tilde{E}_x(-z), \hat{O}_I \tilde{H}_y(z) = \tilde{H}_y(-z)$, respectively, then

$$\begin{aligned} \Gamma_{lj}(-q) &= \int_0^{-\Lambda} [\tilde{E}_x^{(l)*}(-z)\tilde{H}_y^{(j)}(-z) + \tilde{H}_y^{(l)*}(-z)\tilde{E}_x^{(j)}(-z)] d(-z) \\ &= -\int_0^{\Lambda} [\tilde{E}_x^{(l)*}(z')\tilde{H}_y^{(j)}(z') + \tilde{H}_y^{(l)*}(z')\tilde{E}_x^{(j)}(z')] dz' = -\Gamma_{lj}(q). \end{aligned} \quad (28)$$

Thus the Floquet Hamiltonian $\tilde{H}_F$ is asymmetric about $q = 0$, which could lead a nonreciprocal band structure.

In Figs. 7(a) and (7b), we showed the quasienergy bands for $\beta = 0$ and $\beta = \Omega$, respectively, considering the coupling between the positive and positive bands. When $\beta = 0$, the quasienergy bands are reciprocal and band gaps open symmetrically about $q = 0$. However, when $\beta \neq 0$, the bands become nonreciprocal and the gap sizes in the $q < 0$ and $q > 0$ regions are unequal, see Fig. 7(b).

We also investigated the quasienergy bands due to the coupling between the band 1 and the band -1. For $\beta = 5\Omega$ and $\beta = 0.5\Omega$, the quasienergy bands are shown in Figs. 8(a) and 8(b), respectively. Here we used $\Omega\Lambda / c = 0.5$ and $\delta = 0.5$. The quasienergy bands of the both cases are nonreciprocal. And similar to the case of a space-time modulated homogeneous medium [35], band gaps will open for the small modulation speed, see Fig. 8(a), while pairs of EPs will form for the high modulation speed, see Fig. 8(b).

## VI. Conclusions

In summary, we studied the quasienergy bands of the one-dimensional time-Floquet PC which contains two alternating layers with one layer is time-periodically modulated on the permittivity, using the Floquet Hamiltonian derived from the Maxwell's equations based on the time-dependent perturbation theory. When the modulation function is a function of time $t$ only, the quasienergy bands are proved to be reciprocal. For a small modulation strength, we used the 2×2 reduced Hamiltonian to analyze the evolution of the quasienergy bands in the vicinity of the diabolic points formed by bands differing by an order 1. It is shown that when the time-modulation is on the real part of the permittivity, the diabolic point formed by two positive bands is lifted to open a quasienergy gap, while that formed by a positive and a negative bands spawns into a pair of EPs similar to the case of a time-modulated homogeneous media. To the contrary, when the time-modulation is on the imaginary part of the permittivity, the former forms a pair of EPs, while the latter generates a quasienergy gap.

We also investigated the case that $t$ and $z$ are not separable variables in the modulation function. We showed that the quasienergy bands could be nonreciprocal in this case. When the modulated permittivity is pure real, the coupling between the positive and positive bands still results in quasienergy bands. However, the coupling between the positive and negative bands can lead to quasienergy gaps for a small modulation speed and EP pairs for a high modulation speed, which is also similar to the case of a time-modulated homogeneous medium.

**Acknowledgments**: This work is supported by the National Natural Science Foundation of China (NSFC) (No. 11904237, No. 12074267 and No. 11734012), Natural Science Foundation of Guangdong Province (No. 2020A1515010669), and Research and Development Projects in Key Areas of Guangdong Province (No. 2020B010190001).

**Appendix: Formulas of $\Delta_{lj}$ and $\Gamma_{lj}$.**

The photonic band of AB layered lattice in the static case can be calculated using the transfer matrix method as [44]

$$\cos(q\Lambda) = \cos(k_a d_a)\cos(k_b d_b) - \frac{1}{2}\left(\frac{z_a}{z_b} + \frac{z_b}{z_a}\right)\sin(k_a d_a)\sin(k_b d_b), \tag{A1}$$

where $k_i = \sqrt{\varepsilon_i}\,\omega/c, (i = a \text{ or } b)$ is the wavenumber in the $i$th layer, and $z_i = \sqrt{\mu_i / \varepsilon_i}$ is the impedance of the $i$th layer. For nonmagnetic materials ($\mu_i = 1$), the eigenfield is given by [44]

$$E_x(z) = \begin{cases} t_{12}e^{ik_a z} + t_{11}e^{-ik_a z}, & \text{A layer} \\ s_{11}e^{ik_b z} + s_{12}e^{-ik_b z}, & \text{B layer} \end{cases},$$
$$H_y(z) = \begin{cases} n_a(t_{12}e^{ik_a z} - t_{11}e^{-ik_a z}), & \text{A layer} \\ n_b(s_{11}e^{ik_b z} - s_{12}e^{-ik_b z}), & \text{B layer} \end{cases}, \tag{A2}$$

where $n_i = \sqrt{\varepsilon_i}, (i = a, \text{ or } b)$ is the refractive index of the $i$th layer, and

$$t_{11} = e^{iq\Lambda} - e^{ik_a d_a}\left[\cos(k_b d_b) + \frac{i}{2}\left(\frac{n_b}{n_a} + \frac{n_a}{n_b}\right)\sin(k_b d_b)\right],$$
$$t_{12} = e^{-ik_a d_a}\frac{i}{2}\left(\frac{n_b}{n_a} - \frac{n_a}{n_b}\right)\sin(k_b d_b), \tag{A3}$$
$$\begin{pmatrix} s_{11} \\ s_{12} \end{pmatrix} = \begin{pmatrix} e^{ik_b d_b} & e^{-ik_b d_b} \\ e^{ik_b d_b} & -e^{-ik_b d_b} \end{pmatrix}^{-1} \begin{pmatrix} e^{ik_a d_a} & e^{-ik_a d_a} \\ \frac{n_a}{n_b}e^{ik_a d_a} & -\frac{n_a}{n_b}e^{-ik_a d_a} \end{pmatrix} \begin{pmatrix} t_{12} \\ t_{11} \end{pmatrix}.$$

According to the definition,

$$\Delta_{lj} = \langle l | \vec{B}_v j \rangle = \frac{\delta\vartheta}{\sqrt{\iota_l \iota_j}}, \tag{A3}$$

where

$$\vartheta = \int_0^\Lambda E_x^{(l)*} E_x^{(j)*} dz, \ \iota_i = \int_0^\Lambda E_x^{(i)*} \varepsilon(z) E_x^{(i)} dz + \int_0^\Lambda H_y^{(i)*} \mu(z) H_y^{(i)} dz, \tag{A4}$$

with $E_x^{(i)}, H_y^{(i)}$ being the electromagnetic fields of the *i*th static band. Substituting Eq. (A2) into Eq. (A4), we obtain

$$\iota_i = \varepsilon_a (I_a^e + I_a^h) + \varepsilon_b (I_b^e + I_b^h), \tag{A5}$$

where

$$\begin{aligned}
I_a^e &= d_a(|t_{11}^{(i)}|^2 + |t_{12}^{(i)}|^2) + \frac{2}{k_a}\sin(k_a d_a)\,\mathrm{Re}(t_{11}^{(i)*} t_{12}^{(i)} e^{ik_a d_a}), \\
I_b^e &= d_b(|s_{11}^{(i)}|^2 + |s_{12}^{(i)}|^2) + \frac{2}{k_b}\sin(k_b d_b)\,\mathrm{Re}(s_{11}^{(i)} s_{12}^{(i)*} e^{ik_a(d_a+\Lambda)}), \\
I_a^h &= d_a(|t_{11}^{(i)}|^2 + |t_{12}^{(i)}|^2) - \frac{2}{k_a}\sin(k_a d_a)\,\mathrm{Re}(t_{11}^{(i)*} t_{12}^{(i)} e^{ik_a d_a}), \\
I_b^h &= d_b(|s_{11}^{(i)}|^2 + |s_{12}^{(i)}|^2) - \frac{2}{k_b}\sin(k_b d_b)\,\mathrm{Re}(s_{11}^{(i)} s_{12}^{(i)*} e^{ik_a(d_a+\Lambda)}),
\end{aligned} \tag{A6}$$

and

$$\begin{aligned}
\vartheta &= \frac{\tau_1 + \tau_2 + \tau_3}{(k_b^{(j)} - k_b^{(l)})(k_b^{(j)} + k_b^{(l)})}, \\
\tau_1 &= -ik_a^{(l)}(t_{11}^{(j)} + t_{12}^{(j)})(t_{11}^{(l)*} - t_{12}^{(l)*}) - ik_a^{(j)}(t_{11}^{(j)} - t_{12}^{(j)})(t_{11}^{(l)*} + t_{12}^{(l)*}), \\
\tau_2 &= i(k_a^{(l)} + k_a^{(j)})[t_{11}^{(j)} t_{11}^{(l)*} e^{-i(k_a^{(j)} - k_a^{(l)})d_a} - t_{12}^{(j)} t_{12}^{(l)*} e^{i(k_a^{(j)} - k_a^{(l)})d_a}], \\
\tau_3 &= i(k_a^{(j)} - k_a^{(l)})[t_{11}^{(j)} t_{12}^{(l)*} e^{-i(k_a^{(j)} + k_a^{(l)})d_a} - t_{12}^{(j)} t_{11}^{(l)*} e^{i(k_a^{(j)} + k_a^{(l)})d_a}].
\end{aligned} \tag{A7}$$

Using Eq. (A2), we have

$$\varpi_{lj} = \int_0^\Lambda E_x^{(l)*} H_y^{(j)} dz = \frac{1}{(k_a^{(j)} - k_a^{(l)})(k_a^{(j)} + k_a^{(l)})}(\xi_1 + \xi_2 + \xi_3 + \varsigma_1 + \varsigma_2 + \varsigma_3 + \varsigma_4), \tag{A8}$$

where

$$\begin{aligned}
\xi_1 &= in_a(k_a^{(j)} + k_a^{(l)})(t_{11}^{(j)} t_{11}^{(l)*} + t_{12}^{(j)} t_{12}^{(l)*}) + in_a(k_a^{(j)} - k_a^{(l)})(t_{12}^{(j)} t_{11}^{(l)*} + t_{11}^{(j)} t_{12}^{(l)*}), \\
\xi_2 &= -e^{-i(k_a^{(j)} - k_a^{(l)})d_a} in_a(k_a^{(j)} + k_a^{(l)}) t_{11}^{(j)} t_{11}^{(l)*} - e^{i(k_a^{(j)} - k_a^{(l)})d_a} in_a(k_a^{(j)} + k_a^{(l)}) t_{12}^{(j)} t_{12}^{(l)*}, \\
\xi_3 &= -e^{-i(k_a^{(j)} + k_a^{(l)})d_a} in_a(k_a^{(j)} - k_a^{(l)}) t_{11}^{(j)} t_{12}^{(l)*} - e^{i(k_a^{(j)} + k_a^{(l)})d_a} in_a(k_a^{(j)} - k_a^{(l)}) t_{12}^{(j)} t_{11}^{(l)*}, \\
\varsigma_1 &= in_b(k_b^{(j)} + k_b^{(l)}) s_{11}^{(j)} s_{11}^{(l)*}[e^{i(k_b^{(j)} - k_b^{(l)})d_a} - e^{i(k_b^{(j)} - k_b^{(l)})\Lambda}], \\
\varsigma_2 &= in_b(k_b^{(j)} - k_b^{(l)}) s_{12}^{(j)} s_{11}^{(l)*}[e^{-i(k_b^{(j)} + k_b^{(l)})d_a} - e^{-i(k_b^{(j)} + k_b^{(l)})\Lambda}], \\
\varsigma_3 &= in_b(k_b^{(j)} - k_b^{(l)}) s_{11}^{(j)} s_{12}^{(l)*}[e^{i(k_b^{(j)} + k_b^{(l)})d_a} - e^{i(k_b^{(j)} + k_b^{(l)})\Lambda}], \\
\varsigma_4 &= in_b(k_b^{(j)} + k_b^{(l)}) s_{12}^{(j)} s_{12}^{(l)*}[e^{-i(k_b^{(j)} - k_b^{(l)})d_a} - e^{-i(k_b^{(j)} - k_b^{(l)})\Lambda}].
\end{aligned} \tag{A9}$$

Then the matrix element $\Gamma_{lj}$ is obtained as

$$\Gamma_{lj} = \frac{(\varpi_{lj} + \varpi_{jl}^{*})}{\sqrt{\iota_l \iota_j}}. \tag{A10}$$

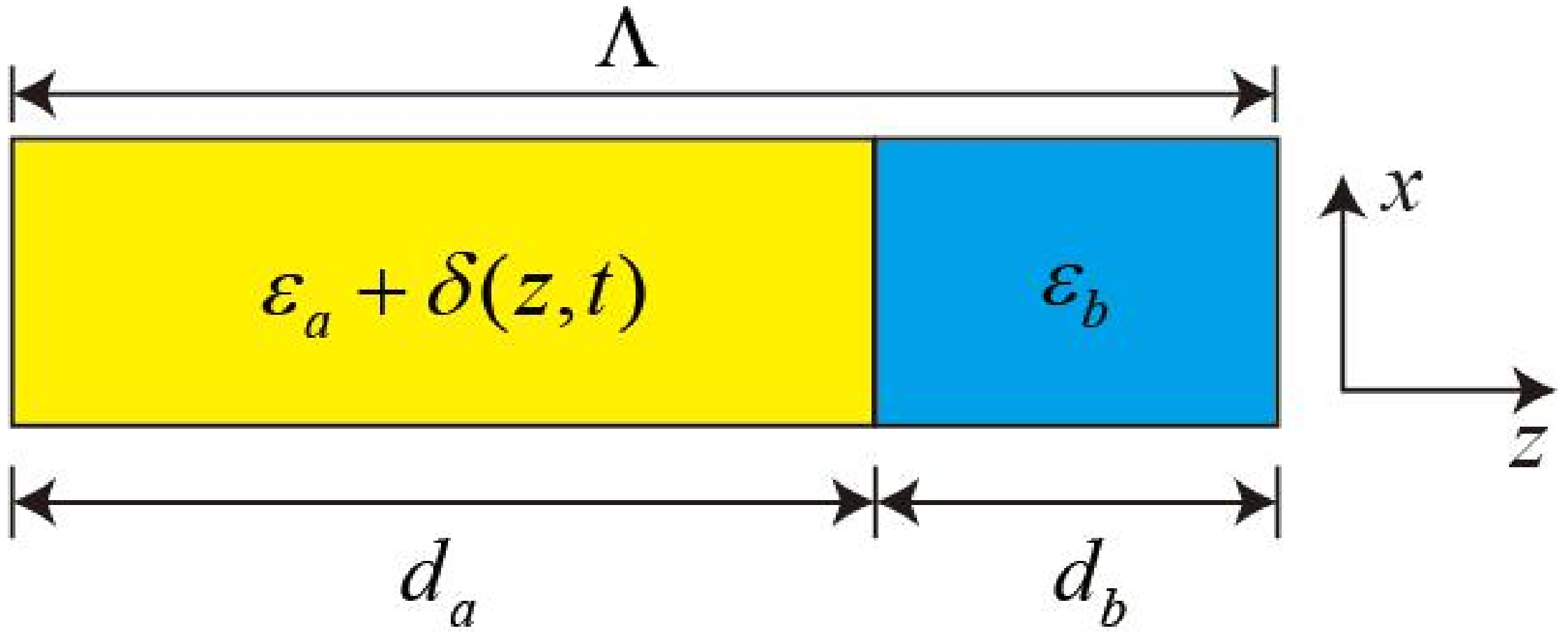

Fig. 1. The unit cell of the 1D time-Floquet PC. The unit cell contains two layers. The first layer (yellow) has a thickness $d_a$ and time-periodic dielectric constant $\varepsilon_a + \delta(z,t)$ , and the second layer (blue) has a thickness $d_b$ and time-independent dielectric constant $\varepsilon_b$.

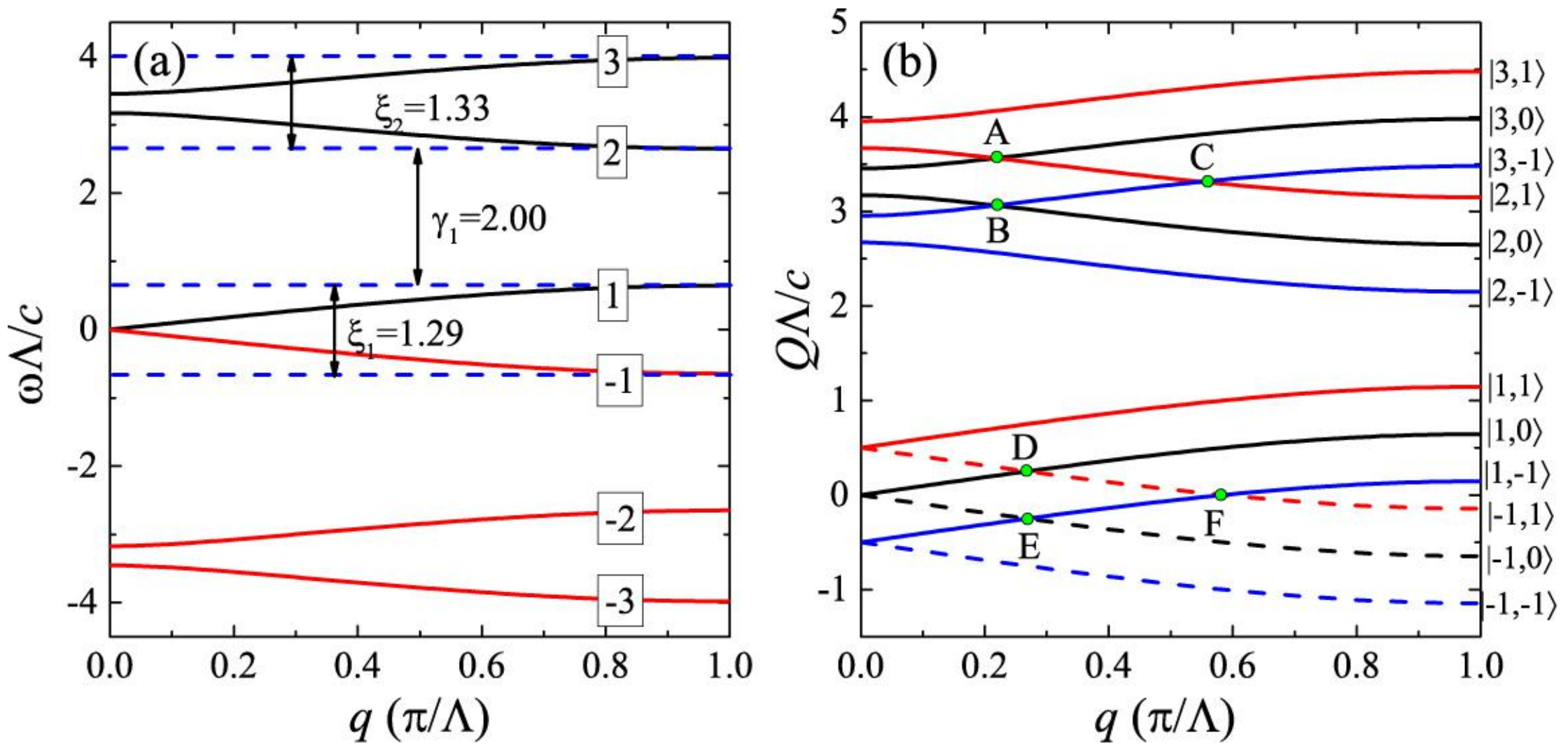

Fig. 2. (a)The band structure in the static case. The positive and negative bands are shown by black and red lines, respectively. The band index is indicated on each band. (b) The quasienergy band in the limit $\delta \to 0$ for $\Omega\Lambda/c = 0.5$ . The 0th, 1st and -1st order bands are shown by black, red and blue lines, respectively. The positive and negative bands are represented by solid and dashed lines, respectively. The green points denote the diabolic points induced by the crossing of bands of different orders.

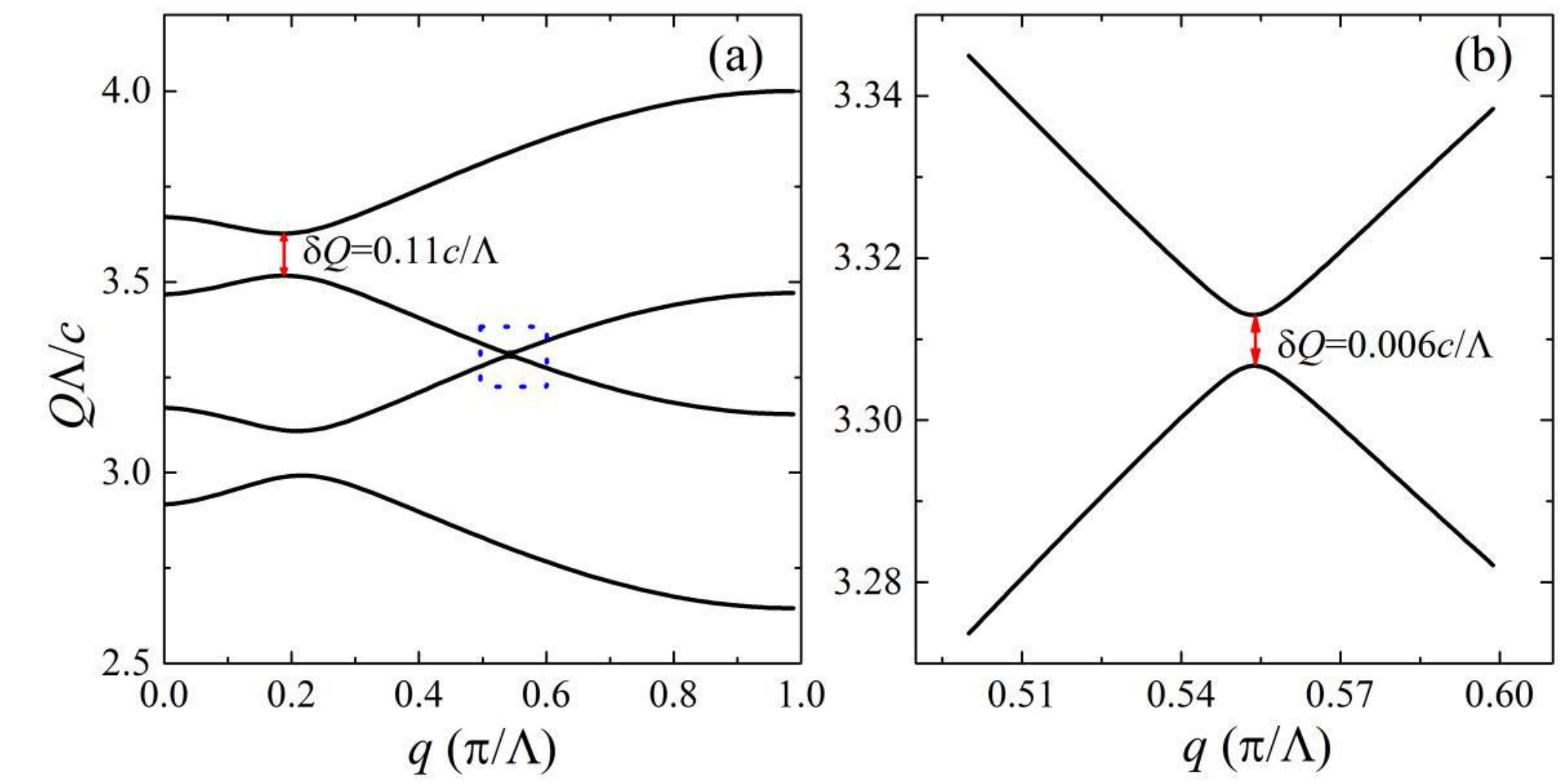

Fig. 3. (a) The quasienergy bands for $\delta = 0.1$ and $\Omega\Lambda/c = 0.5$ . Here, we only showed the bands evolved from $|2,0\rangle, |2,1\rangle, |3,-1\rangle$ and $|3,0\rangle$ . (b) The zoomed-in view of a piece of band in (a) (as marked by the blue dotted rectangle).

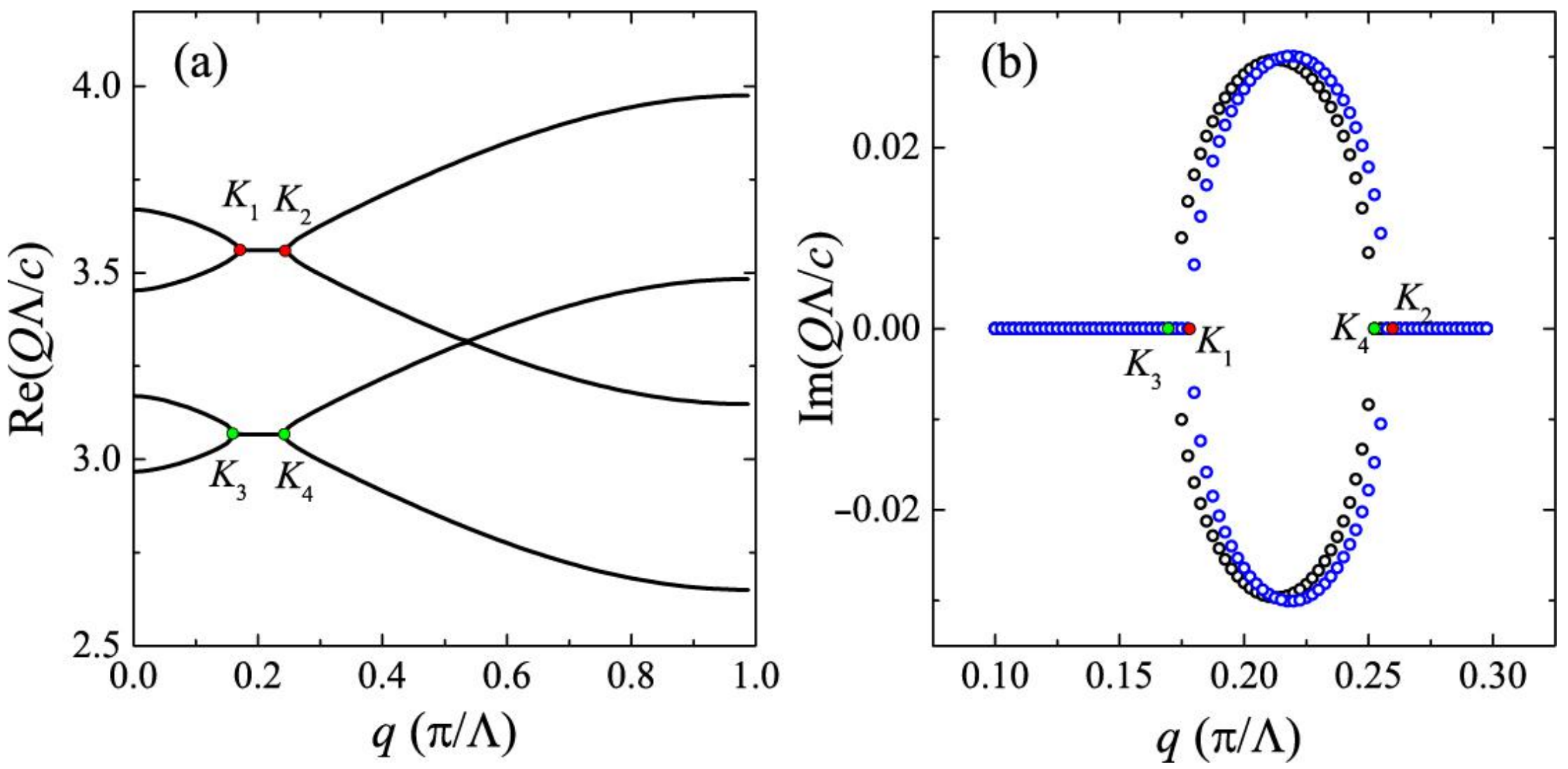

Fig. 4. (a) The real and (b) imaginary parts of the quasienergy bands for $\delta = 0.1i$ and $\Omega\Lambda/c = 0.5$ . The red and green dots denote the two pairs of EPs.

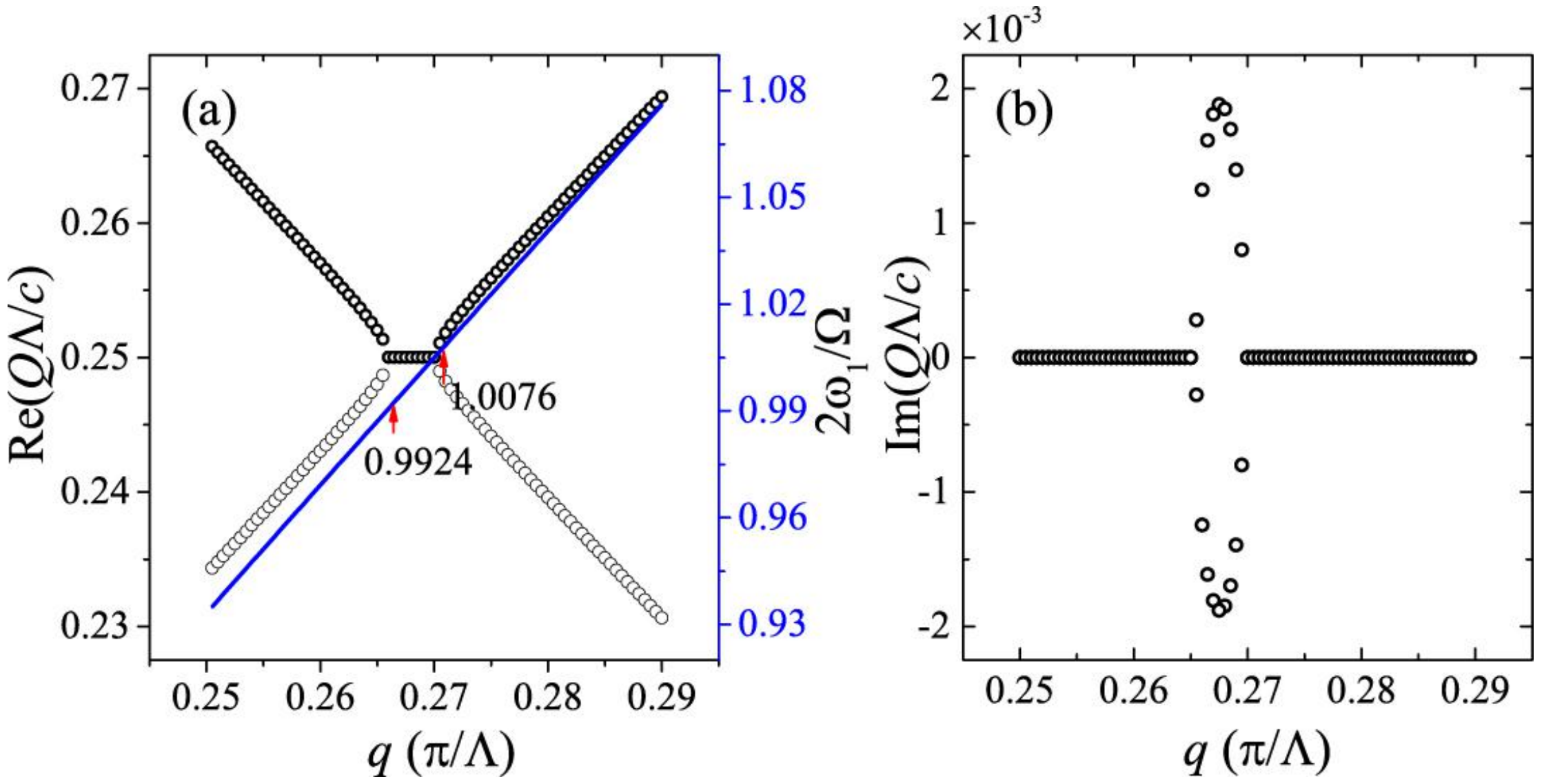

FIg.5. (a) The black circles represent the real part of the quasienergy bands and the blue line denotes the static band $\omega_1$. (b) The imaginary parts of the quasienergy bands. Here $\delta = 0.2, \Omega\Lambda / c = 0.5$.

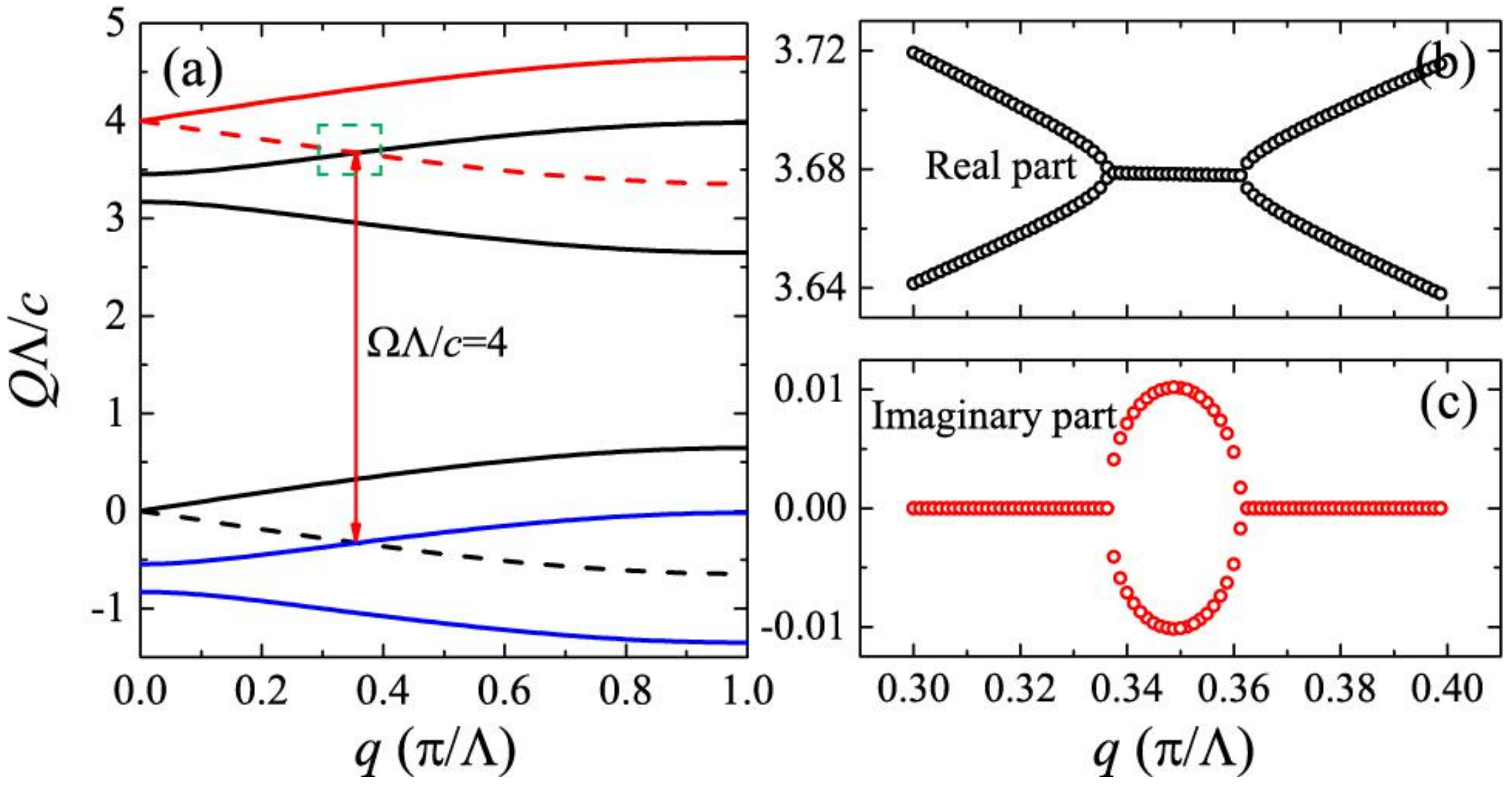

Fig. 6. (a) The quasienergy bands in the limit of $\delta \to 0$ for $\Omega\Lambda/c = 4$. (b) The real and (c) the imaginary parts of a piece of quasienergy bands (marked by the green dashed rectangle in (a)) when $\delta = 0.1$.

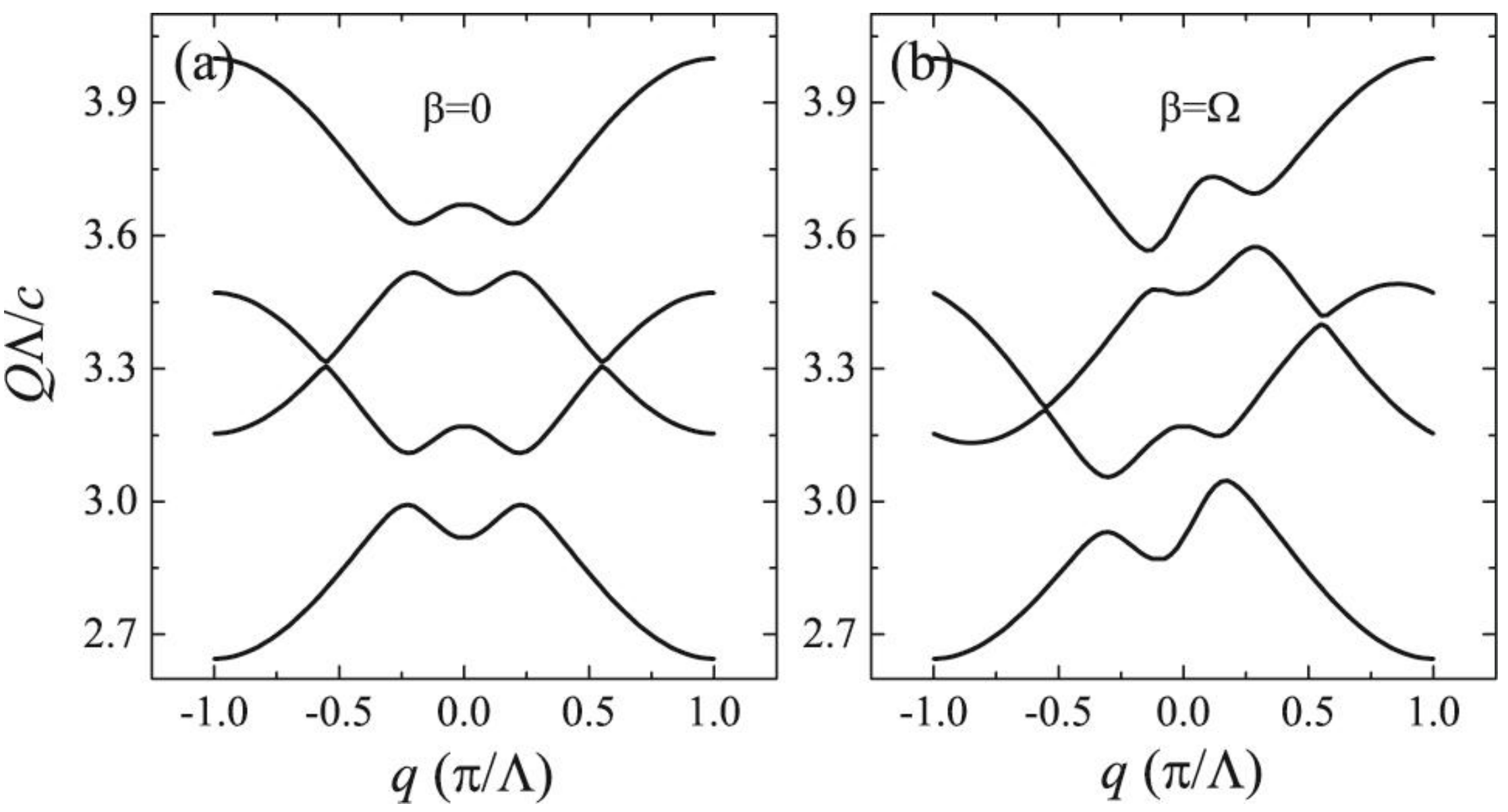

Fig. 7. The quasienergy bands for (a) $\beta = 0$ and (b) $\beta = \Omega$ when $\delta = 0.1$ and $\Omega\Lambda/c = 0.5$.

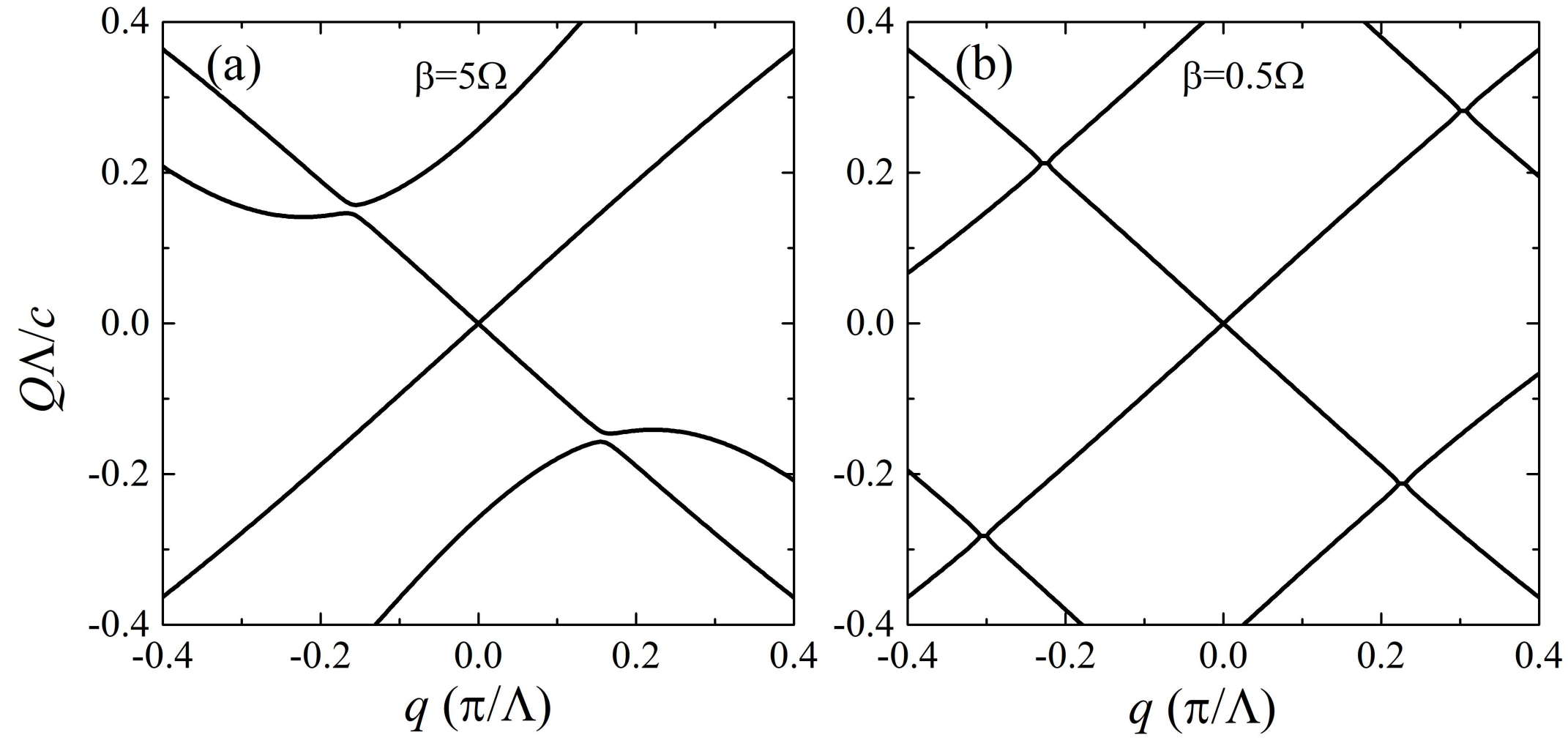

Fig. 8. (a) The quasienergy bands for (a) $\beta = 5\Omega$ and (b) $\beta = 0.5\Omega$. In (b), only the real part of the quasienergy bands is shown. Other parameters: $\Omega\Lambda/c = 0.5, \delta = 0.5$.